\documentstyle[11pt]{article}
\normalbaselines

\def\cK{ {\, \cal K}}
\def\cP{ {\, \cal P}}
\def\HH{{\mbox{${H \! \!  I}$}}}
\def\H{\ifmmode{\HH}\else{\HH\ }\fi}  
\def\and{\quad \mbox{ and } \quad }
\def\Abt{\tilde{\mbox{\boldmath{$A$}}}}
\def\cH{{\cal H}}
\def\crs{commutation relations \ }
\def\gbh{\, \hat{\cal G}}
\def\gbhh{\, \hat{\mbox{$\cal{G}$}}}

\def\pb{\mbox{\boldmath{$p$}}}

\def\Ab{\mbox{\boldmath{$A$}}}
\def\Mb{\mbox{\boldmath{$M$}}}
\def\Jb{\mbox{\boldmath{$J$}}}
\def\Lb{\mbox{\boldmath{$L$}}}
\def\rb{\mbox{\boldmath{$r$}}}
\def\A{\mbox{\boldmath{$A$}}}

\def\L{\mbox{\boldmath{$L$}}}

\def\A{\ifmmode{\Ab}\else{\Ab\ }\fi}  
\def\L{\ifmmode{\Lb}\else{\Lb\ }\fi}
\def\M{\ifmmode{\Mb}\else{\Mb\ }\fi}
\def\gbh{\ifmmode{\gbhh}\else{\gbhh\ }\fi}  %

\def\hH{\hat H}
\newcommand{\be}{\begin{equation}}
\newcommand{\ee}{\end{equation}}
\newcommand{\br}{\begin{eqnarray}}
\newcommand{\er}{\end{eqnarray}}

\newcommand{\nn}{\nonumber}
\newcommand{\for}{\qquad \mbox{ for}\quad}
\newcommand{\aand}{\qquad \mbox{ and}\quad}
\newcommand{\where}{\quad \mbox{ where}\quad}

\def\lb#1{\label{eq:#1}}
\def\rf#1{(\ref{eq:#1})}
\def\x{\times}

\def\ajp#1#2#3{\bibitem{#1} #2, {\it Am. J. Phys.} {\bf #3}}
\def\ijmp#1#2#3{\bibitem{#1} #2, {\it Intern. J. Mod. Phys.} {\bf #3}}
\def\pl#1#2#3{\bibitem{#1} #2, {\it Phys. Lett.} {\bf #3} }

\begin{document}
\bibliographystyle{unsrt}
\hfill{hep-th/9408080}
\vbox{\vspace{1mm}}
\begin{center}
{\LARGE \bf The Dynamical Algebra of the Hydrogen Atom \\
as a Twisted Loop Algebra\footnote{Based on a talk given by
J. Daboul at the XX International
Colloquium on ``Group Theoretical Methods in Physics", Osaka, July 3-9,
1994} }\\[5mm]

Claudia \ Daboul $^\ast$, \underline{Jamil \ Daboul}$\, \sharp$,
Peter \ Slodowy$^\ast$ \\[3mm]
$^\ast${\it  Mathematisches Seminar, Universit\"at Hamburg\\
Bundesstr. 55, D-2000 Hamburg 13, Germany}\\[3mm]
$\sharp${\it Physics Department, Ben Gurion University of the Negev\\
84105 Beer Sheva, Israel (e-mail: daboul@bguvms.bgu.ac.il)}\\[5mm]
\end{center}

\begin{abstract}
\begin{large}
We show that the dynamical symmetry of the hydrogen atom leads
in a natural way to an infinite-dimensional algebra, which we
identify as the positive subalgebras of twisted Kac-Moody algebras
of $ so(4)$. We also generalize our results to the
$N$-dimensional hydrogen atom.
For odd $N$, we identify the dynamical algebra with the positive part of
the twisted algebras $\hat {so}(N+1)^\tau$.
However, for even $N$ this algebra corresponds to
a parabolic subalgebra of the untwisted loop algebra $\hat{so}(N+1)$.

\end{large}
\end{abstract}

\section{Introduction}
It is well known that in the Kepler problem,
defined by the Hamiltonian $H= \pb^2/ 2\mu - \alpha /r$,
the {\bf Runge-Lenz vector}$^{1-2}$
\be \A ={1\over 2}[\pb \x \Lb -\Lb \x \pb]-{\mu  \alpha } {\rb \over r}
 ~, \lb{rv}
\ee
is conserved, $[H,\A]=0$. Here, $\mu$ is the (reduced) mass, and
$\alpha$ is any coupling
constant, which for the Hydrogen atom is equal to $e^2$.

The components of \A and the angular momentum vector \L have
the following \crs:
\be [L^i ,L^j ] = i\hbar \epsilon_{ijk} L^k\; ,\quad
[L^i ,A^j]  = i\hbar  \epsilon_{ijk} A^k\; , \quad
 [A^i,A^j]  = i\hbar  \epsilon_{ijk} (-2\mu H) L^k\; .  \lb{ac} \ee
The 6 operators $L^i$ and $A^i$  do {\em  not} form a closed
finite-dimensional
algebra on the whole Hilbert space $\cH$, because the
Hamiltonian H appears on the r.h.s.\ of \rf{ac}. Therefore, in the
standard treatments one concentrates on individual subspaces $\cH(E)$
which belong to definite
energies $E$. In each such subspace, the Hamiltonian in \rf{ac}
can be replaced by its eigenvalue $E$. This led people to identify
the dynamical algebra with three different algebras, namely
$so(4)$, $e(3)$ and $so(3,1)$, depending on the value of the energy,
as I shall explain later on.
This situation is not satisfactory, since the identification of the
algebra should not depend of the energy. \\

I shall now show that the dynamical algebra of the Kepler problem
can be identified in
a natural way with the infinite dimensional twisted loop algebra of
$so(4)$, and then give a few comments on the  generalization to the
formalism to $N$-dimensional Hydrogen atom.

\section{The standard identification of the dynamical algebra}

Let me first recall that each of the three algebras $so(4)$, $e(3)$ and
$so(3,1)$ is defined in terms of 6 generators, which can be written as
two 3-vectors: $ \Jb$ and $\Mb^\eta$, which obey the following  \crs:
\be [J^i ,J^j ]= i \epsilon_{ijk} J^k\; , \quad
   [J^i, M^{\eta,j}]  = i  \epsilon_{ijk} M^{\eta,k}~,
\quad   [M^{\eta,j},M^{\eta,j}] = \eta i \epsilon_{ijk}
J^k\; ,  \lb{aceta} \ee
where the summation over the repeated index $k$ is implied.
For $\eta=1,0,-1$, the above \crs define the three algebras
$so(4)$, $e(3)$ and $so(3,1)$,
respectively. Note that I am using $J^i$ instead of $L^i$ in the abstract
definition of the algebras \rf{aceta}, in order to distinguish between
the $J^i$ and their differential-operator representations $L^i$.

As I said before, in the usual treatment one concentrates on
individual subspaces $\cH(E)$ which belong to definite
energies $E$. For each such subspace, one can replace the Hamiltonian
in \rf{ac} by its eigenvalue $E$. One then normalizes $A^i$,
and obtains algebras, which are isomorphic to the three given in
\rf{aceta}.

For example, for negative energies, the spectrum is discrete
($E_n$ with $n=1,2,\dots$). Here, one usually defines ``normalized"
Runge-Lenz vectors by, $ \Abt(E_n) :=  \A/\sqrt{-2\mu E_n},$
which lead to \crs similar to those of $so(4)$,
$ [\Abt^i(E_n), \Abt^j(E_n)]= i \hbar \epsilon_{ijk} L^k ~.$
Since the energy subspaces $\cH(E_n)$  have  $n^2$ degenerate levels,
the above procedure leads to $n^2\x n^2$ irreducible matrix representations
of the operators $L^i$ and $\tilde{A}^i(E_n)$ and thus of $so(4)$.

For the positive spectrum, one defines
$ \Abt(E) := \A/\sqrt{2\mu E} $, so that
$ [\Abt^i(E),\Abt^j(E)]= - i \hbar \epsilon_{ijk} L^k $. In this way,
one gets for every $E > 0$ a different representation of $so(3,1)$
in terms of automorphisms (differential operators) on the
subspace $\cH(E)$ .

Finally, for $E=0$ there is no need for normalization, since
the $A^i$ commute among themselves, so that $L^i$ and $A^i$,
when applied to  $\cH(E=0)$ leads automatically to an infinite
dimensional representation of the Euclidean algebra $e(3)$.\\

\section{The Infinite-Dimensional H-algebra \H}

In our new treatment, we keep the operators $A^i$ as they are. Instead,
we include in the algebra all the products of $L^i$ and $A^i$ with
the {\em positive} powers of the Hamiltonian H. Thus, we define
\be
  L^{i}_n := \hH^n L^i \and A^{i}_n := \hH^n A^i\; ,
 \mbox{   where } n\ge 0\; ,\; i=1,2,3,\mbox{  and  }  \hH:=-2\mu H \; .
\lb{heta}\ee
In this way, we obtain a closed but infinite-dimensional algebra,
which we shall call the  {\bf H-algebra} and denote it by $\H$.
The commutation relations follow immediately from \rf{ac}
\be [L^{i}_n, L^{j}_m] =
i\hbar \epsilon_{ijk} L_{n+m}^{k}\; , \quad
    [L^{i}_n, A^{j}_m] =
i\hbar \epsilon_{ijk} A_{n+m}^{k} \;, \quad
    [A^{i}_n, A^{j}_m] =i\hbar \epsilon_{ijk} L_{n+m+1}^{k} \;. \lb{h}\ee
This algebra looks exactly like the loop algebra of $so(4)$, {\em except for
the extra 1 in the lower index of} $L_{n+m+1}^{k}$. Because of this
extra 1, the H-algebra turns out to be isomorphic to the positive part
of the {\em twisted} loop algebra of $so(4)$,
as we shall show below.

\subsection{Quotient algebras of \H: A formal construction}

Even before identifying \H I'll now show how we can formally
reproduce the three algebras
$so(4)$, $so(3,1)$ and $e(3)$, as quotient
algebras of \H, by using the following construction: Clearly,
$ I(c):=(\hH-c) \H $  is an ideal of $\H $
for every real parameter $c$. Therefore, the quotient algebra
$ \H /I(c) $ has only 6 basis elements, which can
be represented by $L^i$ and $A^i$. The elements are the
subspaces $\hat L^i \equiv L^i+ I(c)$ and $\hat A^i \equiv A^i+I(c)$. By
recalling that in the quotient algebra
the ideal $I(c)$ acts as the zero element, we easily get the
following commutation relations:
\be [\hat L^i ,\hat L^j]=i\hbar\epsilon_{ijk}\hat L^k\; ,
\quad  [\hat L^i ,\hat A^j] = i\hbar \epsilon_{ijk} \hat A^k\; ,
\quad  [\hat A^i,\hat A^j] = \hH \; i\hbar \epsilon_{ijk}\hat L^k
= c \; i\hbar \epsilon_{ijk} \hat L^k\; ,  \lb{aac} \ee
which are similar to \rf{aceta}. Therefore, the commutation relations
\rf{aac}
define algebras which are isomorphic to $so(4)$, $so(3,1)$ and $e(3)$,
for $ c>0, c<0$, and $c=0$, respectively. The above construction can be
summarized, as follows:
\be  Q(c)\equiv \H / I(c) \simeq \left\{
\begin{array}{ll}
so(4), & \qquad \for c>0, \\
so(3,1), & \qquad \for c<0, \\
e(3), & \qquad \for c=0 .
\end{array} \right .\ee
The use of the ideal $ (\hH-c) \H  $
is practically equivalent to the usual projection procedure
on the eigenspaces $\cH(E)$, if  $c=-2\mu E$.

\section{ The standard and the twisted Kac-Moody algebras of so(4)}

A short review of the basic notions of the standard
and the twisted affine Kac-Moody algebras was given recently by us$^3$.
For more general expositions I refer to references$^{4,5}$.

Here, I shall only give the  definitions for the specific
loop algebras of $so(4)$ and its twisted counterpart:
The loop algebra of $so(4)$ is obtained by
taking infinitely many copies of the 6 original generators.
These copies are distinguished by a lower index $ n\in Z$. Thus the loop
algebra  is generated by the following set of elements (From now on we
shall use $M^i$ instead of $M^{\eta=1,i}$, which was defined in
Eq. \rf{aceta}):
\be  \gbh:=\{J^{i}_{n}\} \cup \{M^{i}_{n}\}~, \where i=1,2,3, \aand n\in Z
\ee
The commutation relations among these are:
\be [J^i_m, J^i_n]=i\epsilon_{ijk} J_{ m+ n}^{k}\; , \quad
    [J^{i}_{ m}, M^j_n] =i \epsilon_{ijk} M_{ m+ n}^k \;, \quad
    [M^i_m, M^j_n] =i \epsilon_{ijk} J_{n+m}^{k}
 \;. \lb{kma}\ee
It is easy to see that the following subset of $\gbh$
\be  \gbh^\tau:=\{J^{i}_{2n}\} \cup \{M^{i}_{2n+1}\}\subset \gbh~,
 \where i=1,2,3, \aand n\in Z
\ee
form a {\em subalgebra} of $\gbh$:
\be [J^{i}_{2m}, J^{j}_{2n}] =
i  \epsilon_{ijk} J_{2m+2n}^{k}\; , \quad
    [J^{i}_{2m}, M^{j}_{2n+1}] =
i  \epsilon_{ijk} M_{2m+2n+1}^{k} \;, \quad
 [M^{i}_{2m+1}, M^{j}_{2n+1}] =i  \epsilon_{ijk} J_{2(n+m+1)}^{k}
 \;. \lb{tkma}\ee
This subalgebra is called  the {\bf twisted Loop algebra} of $\hat{so}(4)$.
The $\tau$ denotes the involution automorphism, which is needed to define
the twisting. This is explained in Ref.$^3$.

To get the Kac-Moody algebra from the corresponding loop algebra, one
has to modify the above commutation
relation by adding to the right-hand sides terms that are proportional to
$\cK$, which is an operator which commutes with all the generators
$T^a_n$, and is called the {\bf central element}. In our case, the central
term will be identically zero. For this reason the loop algebra is
sometimes called ``centerless Kac-Moody algebra".

\section{Identification of the H-Algebra with Twisted Loop Algebras}

It is easy to check that the following map from the abstract positive
subalgebra $\cP \equiv \hat{so}(4)^\tau_+$  of $\hat{so}(4)^\tau$
onto the H-algebra $\H$ \rf{h},
\be
\varphi:~ \cP \longmapsto \H ~, \where
\varphi(J^i_{2n})= \frac{1}{\hbar} L^i_n~,
\aand \varphi(M^i_{2n+1})=\frac{1}{\hbar} A^i_n \;,
\quad n\ge 0\;, \lb{id}\ee
is a homomorphism. For example,
\br [\varphi(M^i_{2m+1}),\varphi(M^j_{2n+1})] &=&
\frac{1}{\hbar^2}[A_m^i,A^j_n]= i \frac{1}{\hbar}\epsilon_{ijk}
L_{m+n+1}^k \nn \\
&=& i \epsilon_{ijk}
\varphi(J_{2(m+n+1)}^k) =\varphi([M^i_{2m+1},M^j_{2n+1}]) \; ,\lb{kma1}\er
where we used the \crs \rf{h}.

This map defines a representation of $\cP$ in terms of
the dynamical operators $H$, $L^i$ and $A^i$ of the hydrogen atom.
In fact, the map \rf{id} is an {\em isomorphism}
between $\cP$ and $\H$. This is because H has an
infinite number of different eigenvalues, which insures that
the images  of different $T^a_n \in \cP $
are linearly independent.

\section {Conclusions and Outlook}

Since the famous paper of Pauli in 1926 on the energy levels of the H-atom,
numerous papers have
been written on the symmetry of the Hydrogen atom. I believe that we have
now given the first correct identification of the
dynamical algebra of the H-atom.


We also generalized the whole
formalism to the $N$-dimensional hydrogen atom  \cite{nieto}, and found
that for odd $N$ the dynamical algebra is the positive
subalgebra of the twisted algebra, $\hat{so}(N+1)^\tau$, as expected.
However, for even $N$ the twisted algebra can be untwisted, so that
the dynamical algebra is a {\em parabolic subalgebra} of the
(untwisted) loop algebra $\hat{so}(N+1)$.
We hope to publish the details soon.


\end{document}